# Nonvolatile ferroelectric field control of the anomalous Hall effect in BiFeO$_3$/SrRuO$_3$ bilayer

Z.Y. Ren[1], Z. Yuan[2], L.F. Wang[3], F. Shao[1], P. F. Liu[1], J. Teng[1], K. K. Meng[1], X. G. Xu[1], J. Miao[1, a)] and Y. Jiang[1 a)]

[1] *School of Materials Science and Engineering, University of Science and Technology Beijing, Beijing 100083, China*

[2] *Department of Physics, Beijing Normal University, Beijing 100875, China*

[3] *Department of Physics and Astronomy, Seoul National University, Seoul, Republic of Korea*

**ABSTRACT**

In this work, the BiFeO$_3$ (BFO)/SrRuO$_3$ (SRO) heterostructure was fabricated and the anomalous Hall effect (AHE) was investigated the in BFO/SRO. It is found the nonmonotonic anomalous Hall resistivity behavior in BFO/SRO is originated from the inhomogeneous SRO layer instead of the topological Hall effect. It is surprised that the AHE in BFO/SRO structure can be manipulated by ferroelectric polarization of BFO. Moreover, an inhomogeneous phenomenological model has been applied on those structure. Furthermore, the modification of band structure in SRO under ferroelectric polarization was discussed by first principle calculation. The ferroelectric-manipulated AHE suggests a new pathway to realize nonvolatile, reversible and low energy-consuming voltage-controlled spintronic devices

---

[a)] Electronic mail: j.miao@ustb.edu.cn
[a)] Electronic mail: y.jiang@ustb.edu.cn

**Keywords**

Anomalous Hall effect; SrRuO$_3$/BiFeO$_3$ heterostructure; Ferroelectric polarization-manipulated;

In the past decade, electric field control of spin transport has attracted intensive attentions, providing the potential to reduce the power consumption for spintronic devices, such as memory and logic devices.[1-2] Several interesting physical mechanisms in electric-field control of spin transport were reported, including the interfacial strain/stress coupling,[3-4] redox reactions on metal/oxide interface [5-6] and the accumulation of spin current.[7] Multiferroic heterostructures, consisting of ferroelectric and ferromagnetic layers, have significant advantages in electric field control of spin transporting behaviors. The magnetoelectric coupling strength of multiferroic heterostructures is much larger than that of single phase multiferroic materials.[8] Compared with $GdO_x$,[9] $Al_2O_3$,[10] et al, multiferroic heterostructures can achieve high-speed performance, low voltage and nonvolatile devices.[11-12]

$SrRuO_3$ (SRO) is a type of complex oxide perovskite with itinerant ferromagnetism, which has fascinating properties originating from the strong spin-orbit coupling. The anomalous Hall effect (AHE) in SRO have been studied as related to the Berry curvature, implying that the Hall conductivity $\sigma_{xy}$ depends critically on the band structure details of material magnetization, and particularly on band crossings around the Fermi level.[13-15] Moreover , the temperature dependence of Hall resistivity in SRO is not only nonmonotonic but also changeable on sign of AHE at a critical temperature,[16-17] . This closely related to the temperature-dependent band crossings.[18] Most interestingly, the topological Hall effect (THE) have been observed in perovskite heterostructures $SRO/SrIrO_3$[19-20] and $SRO/BaTiO_3$[21], and then attribute to the formation of skyrmions, because a strong Dzyaloshinskii-Moriya interaction can occur on the artificial interfaces, hosting topologically nontrivial spin textures in SRO. Although the exact mechanism of the AHE and the presence of THE on SRO remain debatable, the electron distributions around Fermi energy surely play an important role in SRO. Here, we show that the electron distributions around Fermi energy in SRO can be affected by ferroelectric polarization.

In this work, we reported the nonmonotonic anomalous Hall resistivity ($R_{AHE}$) behavior in BiFeO$_3$/SRO (BFO/SRO) is originated from the inhomogeneous SRO layer instead of the THE. Furthermore, the polarization-field-induced effects on the AHE of BFO/SRO heterostructure was first realized and both the magnitude and the sign of $R_{AHE}$ can be modulated by manipulating the polarization of BFO.BFO is a multiferroic materials which can provide remnant ferroelectric field, this is nonvolatile to modulate the SRO. Moreover, only by one low pulse voltage can we switch the polarization of BFO, so the modulation of AHE by polarization is reversible and low energy consuming. The nonvolatile modulation of AHE is originated from the change of band structure around the Fermi level in SRO under ferroelectric polarization, which is explored by the first principle calculation. Therefore, the ferroelectric-manipulated AHE suggests a new pathway to study the spin transport in multiferroic heterostructures.

The samples of La$_{0.7}$Sr$_{0.3}$MnO$_3$ (10 nm)/BFO (100 nm)/SRO (20 nm) (LSMO/BFO/SRO) were epitaxially grown on SrTiO3(001) [STO (001)] substrates using pulsed laser deposition (PLD), the LSMO (10 nm) is used as a bottom electrode for the LSMO/BFO/SRO structure. The films of SRO and BFO were grown by PLD using a KrF laser with the fluence around 1.5 J/cm$^2$ and 1.0 J/cm$^2$ and a repetition rate of 3 Hz, respectively. Before deposition, the base pressure of the chamber was less than $1\times10^{-4}$ Pa, while during deposition the oxygen partial pressure was 10 Pa and the temperature was 780 °C. In order to get the high-quality film, after deposition, the oxygen pressure was kept at the $10^4$ Pa and then the films were cooled down to room temperature.

Figure 1(a) shows the X-ray diffraction (XRD) ω-2θ scan pattern for the LSMO/BFO/SRO sample, which determines the phase structure of the LSMO, SRO and BFO films. Obviously, LSMO (001), SRO (001) and BFO (001) peaks can be observed clearly, respectively, which demonstrates the uniaxial orientation growth of LSMO, SRO and BFO. The topographic image of LSMO/BFO/SRO was represented by atomic force microscopy (AFM), as shown in the Fig. 1(b). The scan area is 1×1 µm², the root-mean-square roughness is 0.38 nm for LSMO/BFO/SRO, meaning that the film surfaces are atomically-flat. The geometry experimental Hall-bar is shown in Fig. 1(c), and the Hall-bar was prepared by electron beam lithography and Ar⁺ etching, with the size of 20×20 µm². The resistance measurements were performed via a physical property measurement system (PPMS) and the channel current I along the x-axis is constantly 100 µA.

Figure 1(d) shows the current-voltage (*I-V*) characteristics of the LSMO/BFO/SRO trilayers at 80 K. The *I-V* curve shows the leakage current is low, which means the BFO is high-quality and insulated. In Fig. 2(b), temperature dependence of the longitudinal resistivity ($R_{xx}$) is shown, the resistivity of SRO decreases as temperature decreases, which indicates a metallic property. It also indicates that Curie temperatures ($T_C$) in our SRO film (125 K) is lower than that of SRO bulk value (160 K),[22] which is consistent with previously reported value in SRO films(130 K).[23] It's noteworthy that the current, which is through the SRO only, verified by the leakage mechanism. The Hall resistance ($R_{xy}$) was measured by sweeping the magnetic field $H$ perpendicular to the sample's surface, the magnetic field is from –3 T to 3 T, with the temperature from 50 K to 120 K. The $\rho_{xy}$ can be expressed as $\rho_{xy}=\rho_{OHE}+\rho_{AHE}$ where the two terms denote the ordinary and anomalous Hall resistivities, respectively.[24] The ordinary Hall effect (OHE) is described by $\rho_{OHE}=R_0 H$, where $R_0$ and $H$ are the ordinary Hall coefficient and out-of-plane magnetic field, respectively. The AHE is described by $\rho_{AHE}=R_S M$, where $R_S$ and $M$ are the anomalous Hall coefficient and out-of-plane magnetic moment, respectively. To evaluate the $\rho_{AHE}$, the $\rho_{OHE}$ should be extracted, thus the ordinary part by linearly fitting $\rho_{xy}$ can be determined in the higher magnetic field region.

Figure 2(a) shows the magnetic field dependence of the $R_{AHE}$ in the SRO channel form 50 K to 120 K. The curves exhibit the square-shaped hysteresis indicated the $H$-induced reversal of the magnetization, confirming that the magnetization of the SRO layer have good perpendicular magnetic anisotropy (PMA). It is found that the $R_{AHE}$-$H$ curves show humps from 80 K to 100 K. Recently, a similar nonmonotonic $R_{AHE}$-$H$ behavior was reported in SrIrO$_3$/SRO bilayer [19-20] and BaTiO$_3$/SRO bilayer [21] skyrmions systems. They clarified that these humps should be assigned to the THE. In a prototypical magnetic skyrmions system, the transverse Hall resistivity $\rho_{xy}$ can be decomposed into thee term, $\rho_{xy}=\rho_{OHE}+\rho_{AHE}+\rho_{THE}$ where $\rho_{THE}$ is topological Hall resistivity.[19] However, in (Bi,Mn)$_2$Se$_3$ thin film,[25] the positive and negative anomalous Hall resistances were found to coexist, and the Hall resistivity of (Bi,Mn)$_2$Se$_3$ shows the peak and valley. As Gerber [26] reported, the distinct nonmonotonic features in the Hall effect signal arise in heterogeneous ferromagnets when components of the material exhibit the AHE with opposite polarities. Also, Kan [27] reported the THE-like signal in SRO arising from the inhomogeneous magnetoelectric properties of SRO. Thus, the origination of AHE in SRO may be still to be debated.

In order to clarity the origin of the hump in the SRO/BFO heterostructure, we define $R_{hump}$ as the height of the hump with respect to the saturation resistivity. As indicated in Figs. 2(c) and 2(d) summarized the temperature dependence of $R_{AHE}$, $R_{hump}$, and coercive field ($H_C$). As shown in Fig. 2(c), the sign of $R_{AHE}$ is found to change from negative to positive at $T_S$ = 85 K, where $T_S$ is the transition temperature at which the sign of $R_{AHE}$ reverses. This change of $R_S$ is also confirmed from the temperature dependence of the anomalous Hall conductivity $\sigma_{AHE}$ in SRO, and the sign change of $R_S$ may be attributed to the Berry curvature.[18] And the $R_{hump}$ shows a peak around the $T_S$. As shown in Fig. 2(d), the $H_C$ goes down following the decreasing of temperature, exhibits a discontinuity around the $T_S$. It indicates that all the abnormalities are related to the sign of $R_{AHE}$ reverses in SRO.

To further understand the originating of hump in the SRO/BFO, the non-saturation loops of $R_{AHE}$-$H$ were also investigated at 85 K. As shown in Fig. 2(f), the emergence of the humps and the $R_{hump}$ in the positive field are depended on the non-saturation negative field, which hypothesizing that over the SRO film is inhomogeneous.[27] For example, assuming the SRO film is inhomogeneous, the $H_C$ and $T_S$ of SRO would be different in each domain. As a result, the multichannel AHE with positive and negative polarity can coexist in SRO film, which can give rise the hump in $R_{AHE}$-$H$ curve.Therefore, the nonmonotonic $R_{AHE}$-$H$ behavior in BFO/SRO arises from the inhomogeneous SRO. The inhomogeneity may arise from the inhomogeneous thickness of SRO and the inhomogeneity induced by the ferroelectric field of BFO. The ferroelectric field induced-inhomogeneity can be modulated by manipulating the ferroelectric field of BFO.

Next, the ferroelectric-manipulated AHE in BFO/SRO heterostructures were demonstrated. Ferroelectric properties of the BFO film were investigated using piezoresponse force microscopy (PFM) at room temperature. Fig. 3(a) shows the PFM image of LSMO/BFO as initial state, the ferroelectric domain structures and domain walls of BFO are obvious. Figs. 3(b) and (c) show the phase and amplitude of BFO layer with upward and downward polarization under −9 V and 9 V, respectively. The PFM indicates that the polarization of BFO at initial state is upward, which is consistent with previous report.[28] Besides, a clear *P-E* hysteretic behavior in SRO/BFO was observed in both phase and magnitude signal in Fig. 3(d). These results clearly show good ferroelectric properties of the BFO.

As shown in Fig. 4(a), the impulse gate voltage with +/- 9V (upward/downward polarization of BFO) in the Hall-bar was applied in situ by Keithley 4200, then the gate voltage was removed and measured the $R_{AHE}$-$H$. Figs. 4(b)-(e) show the $R_{AHE}$-$H$ curves of initial state, oppositely switched state and reversibly switched state for heterostructures, at 80 K, 85 K, 90 K and 100 K, respectively. It is noted that the as-grown ferroelectric polarization is upward. Furthermore, when ferroelectric polarization is from upward to downward, the shapes of $R_{AHE}$-$H$ curve change and the decrease of $R_{AHE}$ are clearly observed. Moreover, when ferroelectric polarization is switched back to the original state, the shapes of $R_{AHE}$-$H$ curve are indeed switched back to the initial state. It should be noted that the voltage is removed during the measurement. Thus, the current-induced heating is negligible. Since heating effect would be symmetry of the $R_{AHE}$-$H$ variation with respect to the applied voltage. Noting that the $R_{AHE}$-$H$ are absolutely different for the + 9 V and the – 9 V, but primarily same for the – 9 V and the as-grown state. Thus, the mock effect induced by the Joule heating can be ruled out. Similarly, the strain effect in SRO/BFO heterostructure can also be excluded, since the piezoelectricity in BFO is small and ferroelastic effects would give an even contribution. These results confirm that the AHE in SRO can be modulated by the ferroelectric remnant polarization in BFO along the out-of-plane direction. Especially, this modulation effect on AHE in SRO/BFO is nonvolatile and reversible.

To clearly observe the ferroelectric-manipulated AHE in LSMO/BFO/SRO heterostructures, we summarized the temperature dependence of $R_{AHE}$ and $R_{hump}$. As shown in Fig. 4(f), a vertical translation of $R_{AHE}$ in $R_{AHE}$ ($T$) between the upward and downward polarization can be observed. It is interesting to note that the ferroelectric polarization modulation can give rise to the sign inversion of $R_{AHE}$. As shown in Fig. 4(g), first the $R_{hump}$ of upward polarization is larger than that of downward polarization then the $R_{hump}$ of upward polarization is smaller than that of downward polarization, indicating that a horizontal translation of $R_{hump}$ in $R_{hump}(T)$ between the two ferroelectric polarization states. The vertical translation of $R_{AHE}$ and the horizontal translation of $R_{hump}$ indicate that the $T_S$ in SRO/BFO is 85 K and 86.5 K for the upward and downward polarization, respectively. It is noted that the $R_{AHE}(T)$ ($R_{hump}(T)$) with the upward polarization and with as grown state are near same.

In this work, the modulation depends on the remnant ferroelectric polarization of BFO, thus the modulation can be realized by a low impulse voltage. A nonvolatile and low energy consumption characterizes is important for a voltage-controlled device. As shown in Fig. 5, the AHE under ferroelectric polarization of BFO layer was repeatedly poled upward and downward at 85 K. Thus, ferroelectric polarization-manipulated AHE is reversible and nonvolatile.

Now, we are trying to understand the mechanism of ferroelectric-manipulated AHE in BFO/SRO heterostructures. According to the formulation of a model considering inhomogeneous magnetoelectric properties in SRO,[27] the ferroelectric - manipulated $R_{AHE}$-$H$ curves can be completely reproduced.

$$f(T',H) = \rho_H(T')\{1 - 2H_{Heav}(H - H_C(T')\}g(T') \qquad (1)$$

As Eq. (1) describes, each domain contributes an effective field response to the transverse resistivity. As shown in Fig. 2(c) and Fig. 2(d), the $R_{AHE}(T')$ and $H_C(T')$ are actual temperature dependence of $R_{AHE}$ and $H_C$, which are fitted by linear and quadratic, respectively. $R_{AHE}(T')$=-0.94062+0.01106$T'$ and $H_C(T')$=1.0597($T'$-125)$^2$. The magnetic moment reversal is described by the Heaviside step function $H_{\text{Heav}}(x)$, and $g(T')$ is the Gaussian function describing a distribution of the domain as

$$g(T') = \frac{1}{\sqrt{2\pi T_\sigma^2}} exp\left[\frac{(T'-T)^2}{2T_\sigma^2}\right] \quad (2)$$

$$\Gamma(H) = \int_0^\infty f(T')dT' \quad (3)$$

In Fig. 6(a) and Fig. 6(b), the calculation of the $R_{\text{AHE}}$-$H$ loops by integrating Eq. (3) for upward and downward state are shown, respectively. Both of them are well reproduced in the experiment, indicating that the $R_{\text{AHE}}$-$H$ nonmonotonic behavior in BFO/SRO is arise from the inhomogeneity of SRO film. As shown in Figs. 6(c) and 6(d), the temperature dependence of $R_{\text{AHE}}$ and $R_{\text{hump}}$ are extracted from the calculated loops. Those data show a vertical translation of $R_{\text{AHE}}$ in $R_{\text{AHE}}$ (T) and a horizontal translation of $R_{\text{hump}}$ in $R_{\text{hump}}(T)$ between the two ferroelectric polarization states, which are consistent with the experiment of $R_{\text{AHE}}(T)$ and $R_{\text{hump}}(T)$ in BFO/SRO under the upward and downward polarization.

Neither the skew-scattering theory nor the side-jump theory seems to be adequate to explain the origin of AHE in SRO layer.[18] Recently, the Berry-phase theory has been applied to account for it.[18,28] The first-principles calculation shows that the AHE conductivity fluctuates strongly with the variation of energy, which is related to the Berry curvature. Therefore, in this work, the modulation of AHE in SRO may be related to the change of Berry curvature. In order to further account for the microscopic nature of the ferroelectric-manipulated AHE in BFO/SRO heterostructures, the electronic structure of SRO was calculated using the first-principles calculation. The calculations were performed for the SRO using the Vienna ab initio simulation package (VASP) with generalized gradient approximation.[29] The cutoff energy for the plane-wave basis set was 500 eV. We set the 3D Brillouin zone (BZ) by 9×9×9 k mesh for the primitive cell of SRO and the spin-orbit coupling (SOC) is included in all the calculations. As shown in Fig. 7(a), we assumed that the Ru and O is diverged, described as $\delta_{Ru-O}$ on the basis of ferroelectric proximity effect,[21, 30-32]. The band structure of SRO with $\delta_{Ru-O}$=0 c, 0.01 c, 0.02 c and 0.05 c are shown in Figs. 7(b)-(e), respectively, c=3.99 Å is the lattice constant of SRO. It is found that the band structures of SRO show the difference near the Fermi energy under the different $\delta_{Ru-O}$.[21,32] The Berry curvature is determined by the band structure near the Fermi energy, which is close related the AHE. Furthermore, the $\delta_{Ru-O}$ gradually decrease from the heterointerface to the interior of SRO, which can give rise to inhomogeneous of SRO.[21] Moreover, the $\delta_{Ru-O}$ under the upward and downward polarization are always different,[21,29] thus the modulation of AHE in SRO can be described by the change of Berry curvature to an opposite polarization state.

In summary, the AHE in BFO/SRO heterostructure was investigated. The nonmonotonic $R_{AHE}$-$H$ in BFO/SRO is unrelated to the THE, which is originated from the inhomogeneous crystal structure induced by the ferroelectric field and inhomogeneous thickness of SRO layer. Besides, the nonmonotonic $R_{AHE}$-$H$ in BFO/SRO can be well explained by the inhomogeneous phenomenological model. Furthermore, the AHE in BFO/SRO can be manipulated by the ferroelectric polarization of BFO. That is, the upward and downward polarization field of BFO can increase and decrease the anomalous Hall resistivity, respectively. Moreover, the ferroelectric polarization-manipulated AHE is reversible and nonvolatile. The first-principles calculations show that the modulation of AHE may be attributed to change of the electron distributions around Fermi energy in SRO with the different ferroelectric polarization of BFO The ferroelectric-manipulated AHE suggests a new pathway to realize an nonvolatile, reversible and low energy-consuming in a voltage-controlled spintronic devices

**FIGURE LEGENDS**

*Fig. 1.* (a) XRD pattern and (b) AFM image of the STO/LSMO (10 nm)/BFO (100 nm)/SRO (20 nm) sample. (c) Schematic diagram of the Hall bar structure. (d) *I-V* curves for LSMO (10 nm)/BFO (100 nm)/SRO (20 nm) at 80K.

*Fig. 2.* (a) Magnetic field dependence of $R_{AHE}$ in different temperature, (b) $R_{xx}$, (c) $R_{AHE}$ and $R_{hump}$, (d) $H_C$ as a function of temperature, (e) minor loops of $R_{AHE}$ at 85 K for BFO/SRO. Temperature dependence of $R_{AHE}$ and $H_C$ are fitted by linear and quadratic, respectively (red line).

*Fig. 3.* PFM images of the STO/LSMO (10 nm)/BFO (100 nm) structure: (a) initial phase, and (b) phase and (c) amplitude after written under ± 9 V, respectively. (d) Ferroelectric phase hysteresis and strain loops.

*Fig. 4.* (a) In BFO/SRO heterostructure the ferroelectric polarization as-grown is upward, when with +9 V is downward and when with -9 V is upward. $R_{AHE}$-*H* hysteresis loops of BFO/SRO heterostructure in as-grown state, oppositely switched state and reversibly switched state at (b) 80 K, (c) 85 K, (d) 90 K, (e) 100 K. The arrows show the direction of polarization. (f) $R_{AHE}$ and (g) $R_{hump}$ as a function of temperature under as-grown state, oppositely switched state and reversibly switched state.

*Fig. 5.* The polarization of BFO layer was repeatedly poled upward and downward, the $R_{AHE}$ under upward and downward polarization.

*Fig. 6.* $R_{AHE}$-*H* hysteresis loops reproduced by numerical model with (a) upward, (b) downward polarization in BFO/SRO. Temperature dependence of the calculated (c)

$R_{AHE}$ and (d) $R_{hump}$ with upward and downward.

*Fig. 7.* (a) Schematic diagram of SRO, the ionic displacement between oxygen and Ru along the [001] axis is denoted as $\delta_{Ru-O}$. The band structure of SRO with $\delta_{Ru-O}$ as (b) 0.00 c, (c) 0.02 c, (d) 0.02 c, and (e) 0.05 c.

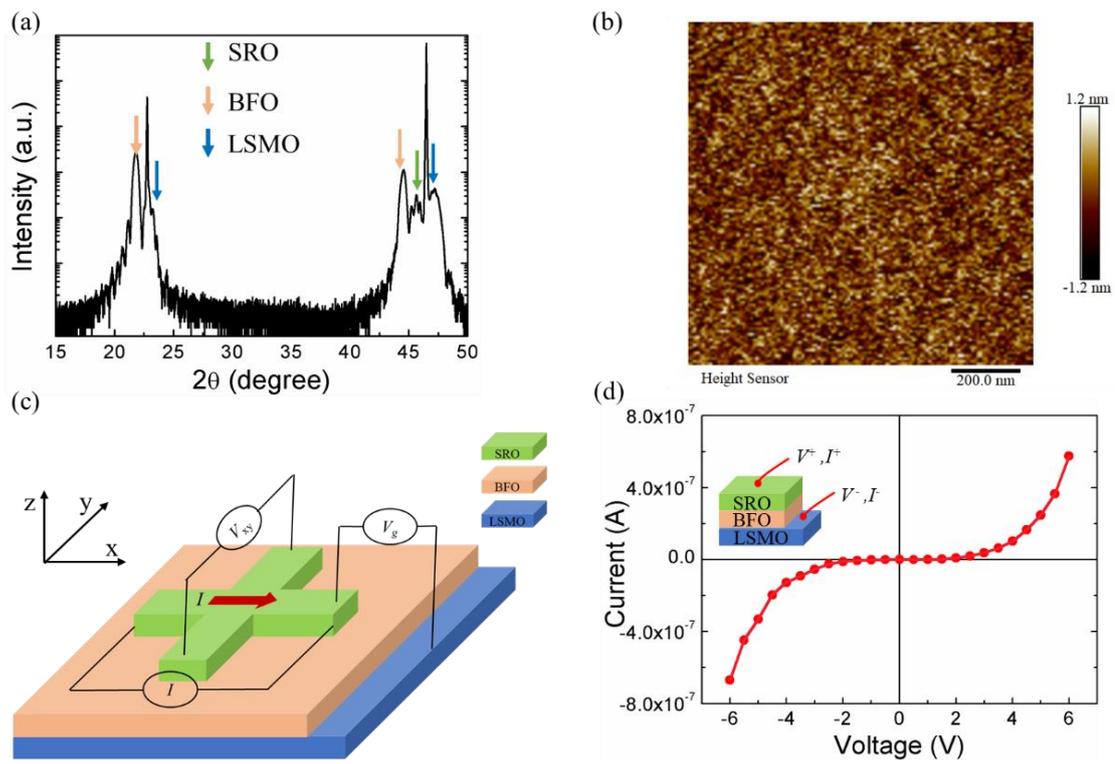

*Figure 1*

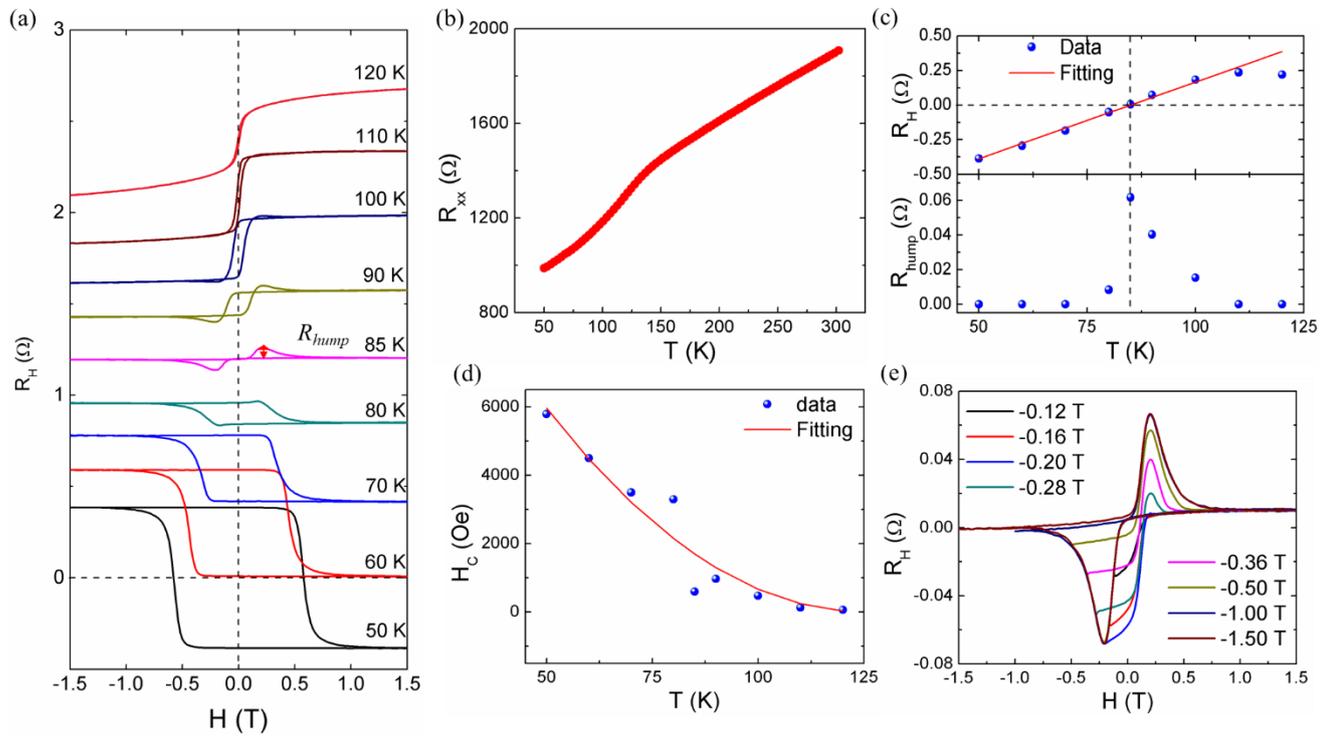

*Figure 2*

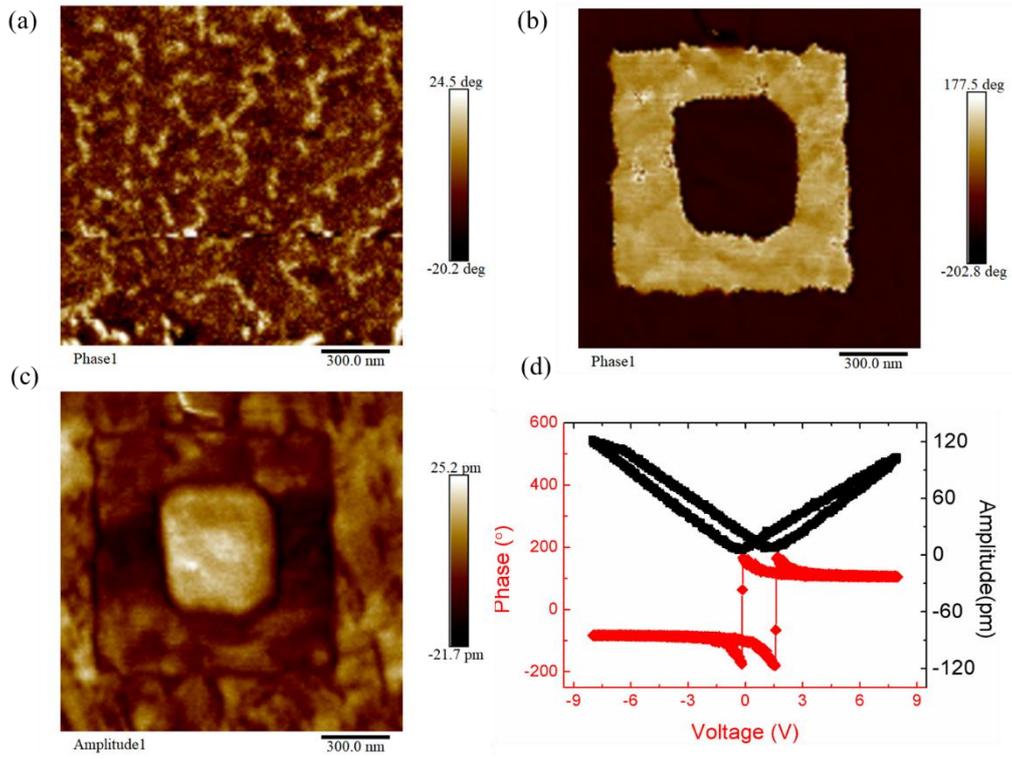

*Figure 3*

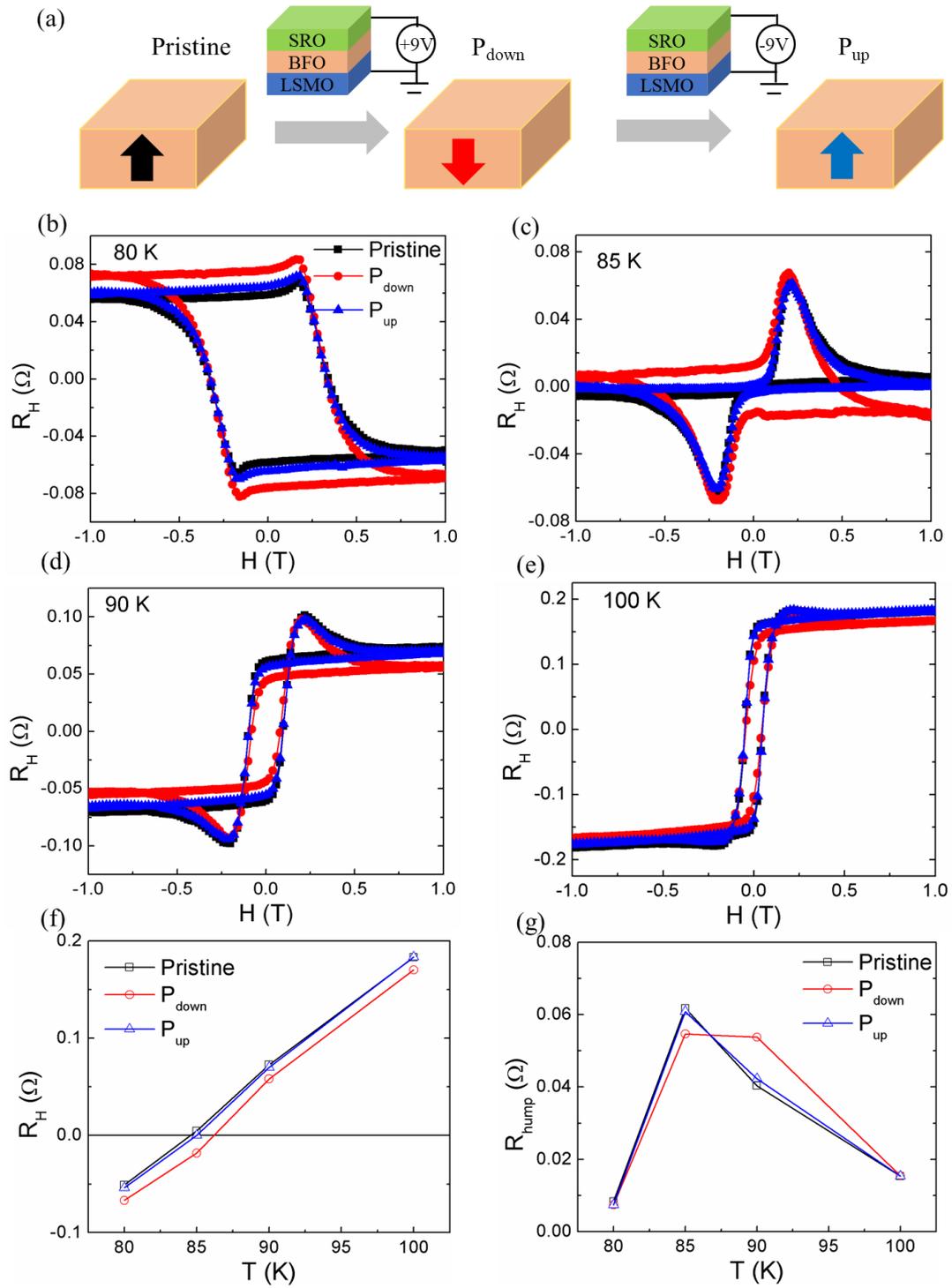

*Figure 4*

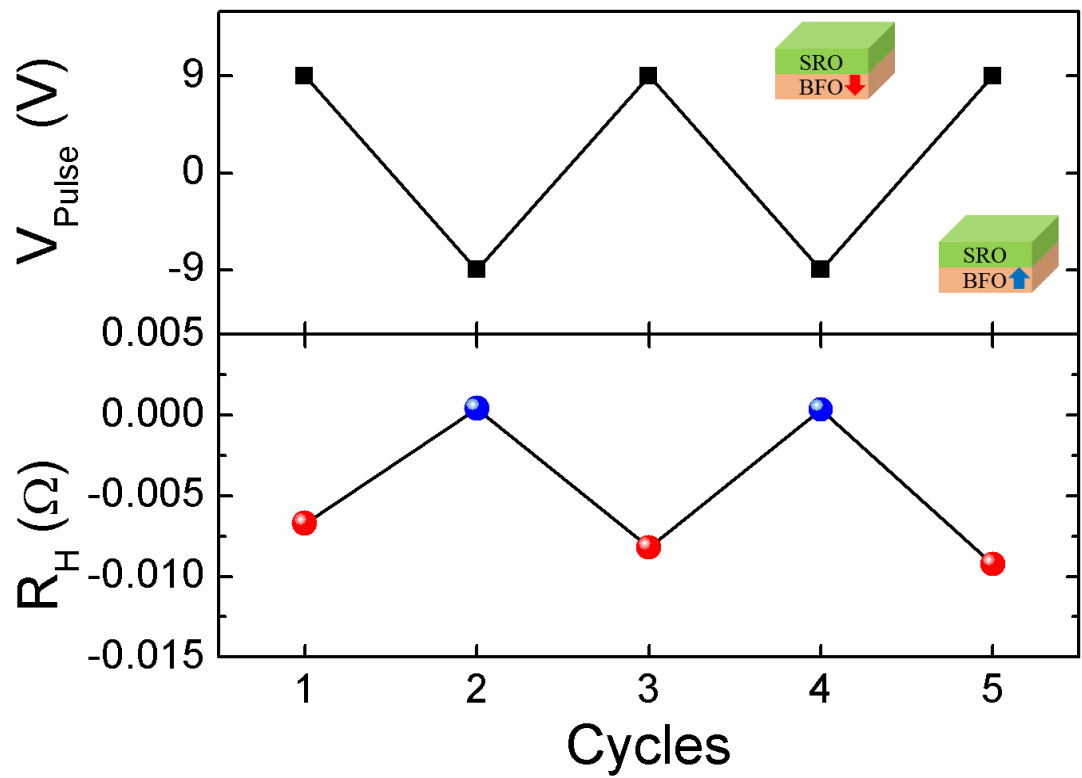

*Figure 5*

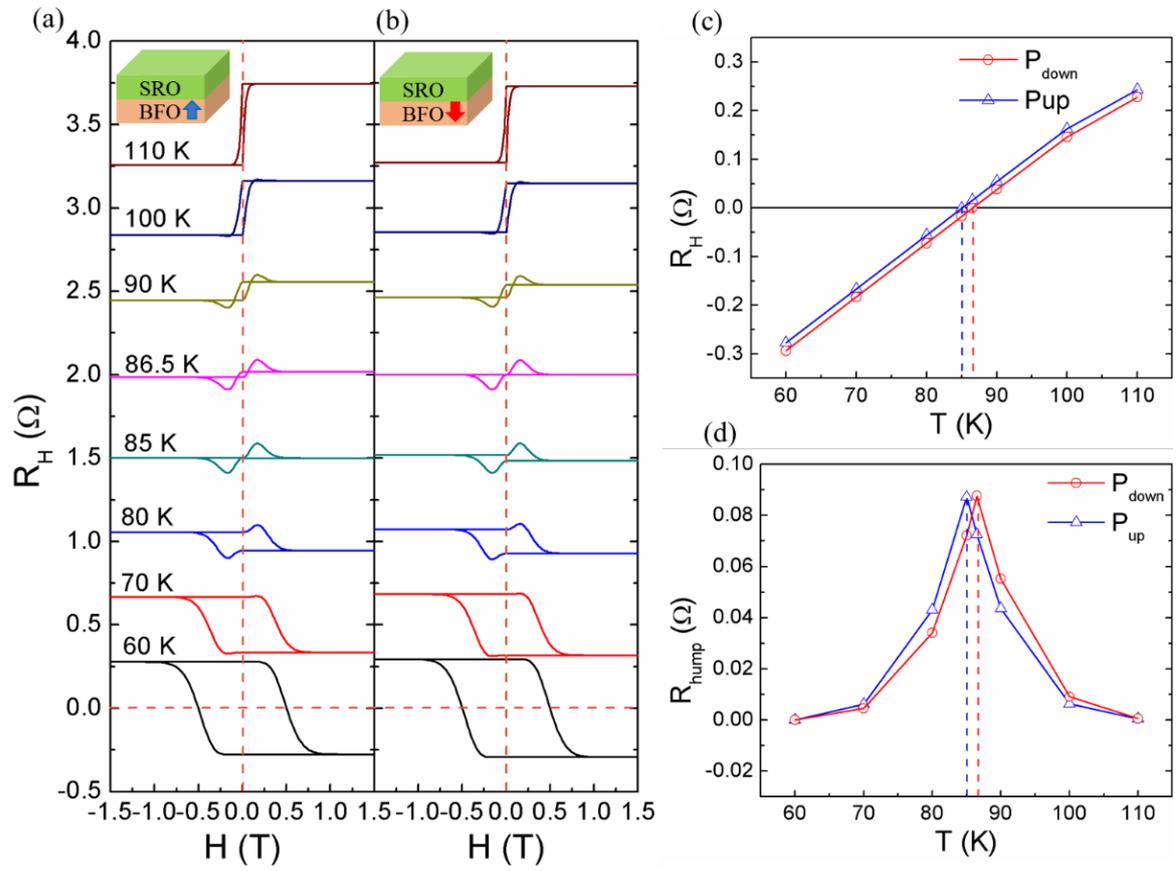

*Figure 6*

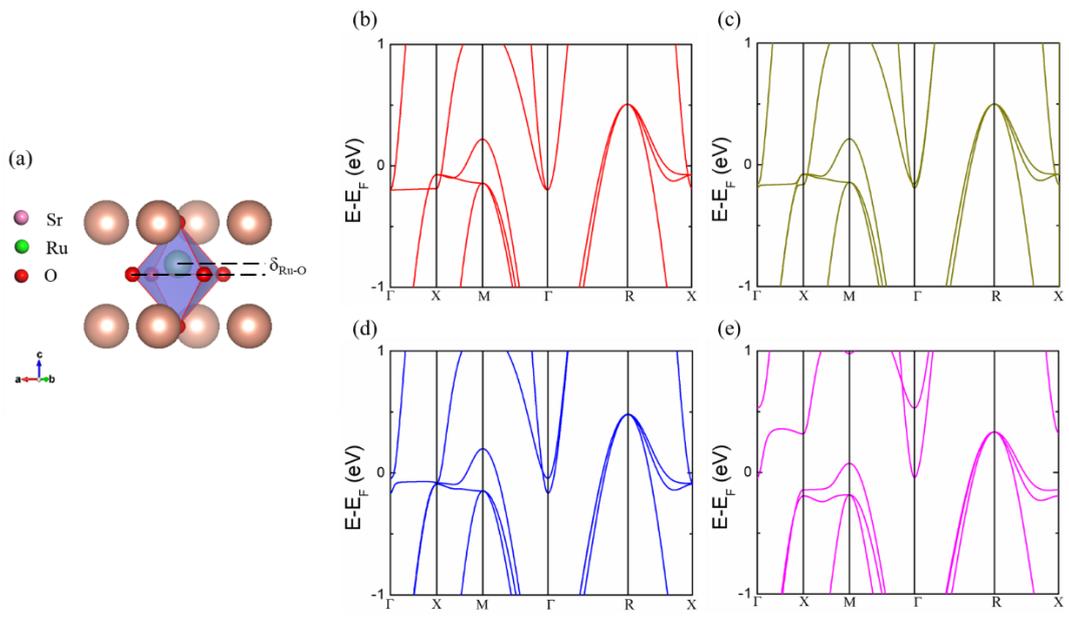

*Figure 7*